# Towards Hardware Implementation of Double-Layer Perceptron Based on Metal-Oxide Memristive Nanostructures

A.N. Mikhaylov, O.A. Morozov, P.E. Ovchinnikov, I.N. Antonov, A.I. Belov, D.S. Korolev, M.N. Koryazhkina, A.N. Sharapov, E.G. Gryaznov, O.N. Gorshkov, V.B. Kazantsev

*Abstract*—Construction and training principles have been proposed and tested for an artificial neural network based on metal-oxide thin-film nanostructures possessing bipolar resistive switching (memristive) effect. Experimental electronic circuit of neural network is implemented as a double-layer perceptron with a weight matrix composed of 32 memristive devices. The network training algorithm takes into account technological variations of the parameters of memristive nanostructures. Despite the limited size of weight matrix the developed neural network model is well scalable and capable of solving nonlinear classification problems.

*Index Terms*—Metal-oxide nanostructure, resistive switching, adaptive behavior, memristive device, artificial neural network, neuromorphic system

## I. Introduction

NEW round of development of brain-inspired neuromorphic systems [1] is one of the breakthrough research trends leading to the development of technological base for the products of an entirely new technical level and new markets [2]. Such products include neuromorphic computing systems [3] and, what is most challenging, neurohybrid systems and technologies based on interfacing electronic neural networks and biological tissues / cultures [4,5] for the development of robotics with elements of artificial intelligence and new methods of diagnosis / therapy and "prosthesis" (partial replacement) of nervous system.

One of the most promising approaches to the hardware implementation of compact and energy-efficient neuromorphic systems relies on the possibility of constructing simple electronic models of neural networks on the base of thin-film memristive nanostructures / devices [6]. Memristive device is a physical model of ideal memristor – a circuit element capable of changing the resistance depending on the input electrical signal [7]. In the case of inorganic solid state memristive devices, the resistance changes due to a reversible transformation of atomic structure in the nanometric region of dielectric or semiconductor (ionic conductor) film deposited between two conductive electrodes. The changed resistive state can be stored in a nonvolatile manner, which allows using memristive nanostructures in the RRAM (Resistive Random Access Memory) and "logic-in-memory" devices [8,9]. Capacitor-like memristive nanostructures based on the combination of metal and oxide thin films (often called as metal-oxide memristive nanostructures) are deposited on conventional silicon substrate and compatible with the modern CMOS fabrication technology of analog and digital electronic circuits. Local reduction-oxidation (redox) processes responsible for the resistive switching phenomenon provide a high degree of miniaturization (down to 10 nm), high speed (switching time as low as 100 ns) and low power consumption (as low as pJ) [10,11].

The ability of memristive device to gradually change the conductivity in response to electrical signals makes it an electronic counterpart of the synapse (synaptic connection between neurons) [12-14]. It follows from the reviews and original reports [15-18] that a considerable progress has been made to date in the simulation of properties and functions of biological synapse by using memristive devices: a continuum of resistive states, synaptic plasticity (short-term and long-term potentiation / depression, STDP – Spike Timing Dependent Plasticity) revealed in the dependence of memristor conductivity on the amplitude, duration, frequency and timing of electrical pulses.

Important issues, when studying the plasticity of memristive devices, are the dynamic range of resistance and its gradual change in the processes of increase (SET) or decrease (RESET) of the conductivity that depend on the programming technique [19]. For example, in contrast to the gradual RESET transition, strong current jumps in the SET process in response to pulses with amplitudes above the threshold provide a wide resistance range, but lead to the necessity of using current compliance. Such peculiarities of a memristive device obviously must depend on specific materials used in its thin-film structure. As it has been shown previously [20], the Au/Zr/oxide/TiN/Ti memristive nanostructures based on different oxide dielectrics exhibit qualitatively similar bipolar resistive switching behavior. However, the substantial

Manuscript received July 27, 2017. This work was supported by the grant of Russian Science Foundation (No. 16-19-00144).

A.N. Mikhaylov, O.A. Morozov, P.E. Ovchinnikov, I.N. Antonov, A.I. Belov, D.S. Korolev, M.N. Koryazhkina, A.N. Sharapov, E.G. Gryaznov, O.N. Gorshkov, V.B. Kazantsev are with the Lobachevsky University, 603950 Nizhny Novgorod, Russia (e-mail: mian@nifti.unn.ru, oa_morozov@nifti.unn.ru, ope@nifti.unn.ru, ivant@nifti.unn.ru, belov@nifti.unn.ru, dmkorolev@phys.unn.ru, mahavenok@mail.ru, asharapov@itc.unn.ru, gryaznov_eg@mail.ru, gorshkov@nifti.unn.ru, kazantsev@neuro.nnov.ru).

difference is revealed in the origin of resistive states in these materials and mechanisms of microscopic processes responsible for the change of resistance. The dynamic resistance range should also depend on the structure and composition of oxide material adding an extra degree of freedom when tuning the adaptive behavior of memristive devices required for their application in neuromorphic systems.

Despite the evident progress in simulation of synaptic plasticity by using memristive devices, fully hardware implementation of a large-scale memristor-based artificial neural network (ANN) is still a great challenge. Previously, the ANN design principles and performance were demonstrated by using digital electronic or software simulators of memristors (e.g. [21-22]). Only in the past two years, first reports were published on successful creation of ANNs based on a limited number of memristive devices taken as connections between the artificial neurons represented by external electronic devices or software modules. Such ANNs include a reconfigurable Hopfield neural network on the array of six independent $HfO_2$-based memristors with active Ni electrodes [23], a single-layer perceptron based on memristive nanocrossbar with 12 × 12 devices of the $Pt/Ti/TiO_x/Al_2O_3/Pt/Ta$ type [24], single- and multi-layer perceptrons based on organic PANI memristors capable to solve both linear and nonlinear problems [25,26]. An interesting report [27] demonstrates the unsupervised learning effect in a probabilistic neural network of the «winner-take-all» (WTA) type using four $Pt/TiO_2/Pt$ memristive devices. It should be mentioned that the above mentioned ANNs on the basis of real metal-oxide memristive devices were capable of solving simple linear problems, whereas more complex functionalities and features of large-scale networks were shown using the software simulators, which reproduce the behavior and variation of parameters of real memristive devices. A large-scale integrated 1T1R array of 1024 metal-oxide memristive devices has also been recently fabricated and tested as synaptic matrix in the single-layer network organized by using software neurons and external equipment [28], but the full chip with hardware neurons and all the peripheral circuits has not been realized yet [29].

Our work focuses on the hardware implementation of a *one-board* ANN prototype in the form of a double-layer perceptron purposed to solve *nonlinear classification problems*. The developed ANN architecture is rather similar to one reported in [26] for organic memristive devices, but our solution employs the array of 32 metal-oxide memristive devices as a weight matrix. The devices are based on the Au/Zr/oxide/TiN/Ti thin-film nanostructures, which are technologically adapted and CMOS-compatible.

## II. STRUCTURE AND PROPERTIES OF MEMRISTIVE DEVICES

Oxide layer in the studied Au/Zr/oxide/TiN/Ti thin-film nanostructure is represented by the $ZrO_2(Y)$ or $SiO_2$ film of various thickness. The first dielectric material is a transition metal oxide with predominantly ionic nature of chemical bond (also known as a solid-state electrolyte), in which the yttrium oxide (12 mol.%) doping agent stabilizes the cubic phase of zirconia and determines certain concentration of oxygen vacancies playing a key role in resistive switching – formation and local destruction of conductive channels (filaments) in oxide film [30-32]. The second kind of oxide is a "native" dielectric for CMOS-technology and is also known as intrinsic resistive switching material [33]. In contrast to $ZrO_2(Y)$, the conductive channels in $SiO_2$ are grown due to the breaking of Si-O-Si bond under electric stress followed by the formation of neutral oxygen vacancy (Si-Si bond as a constituent of silicon filament) and oxygen interstitial atoms (ions), migration of which provides the filament oxidation and recovery [33-34].

Memristive nanostructures schematically illustrated in Fig. 1 were deposited on the oxidized silicon substrates by the method of magnetron sputtering. The switching dielectric layer is a $ZrO_2(Y)$ or $SiO_2$ film with a thickness of 40 or 60 nm, upper electrode – Au film (40 nm) with Zr adhesion sublayer (3 nm), bottom electrode – TiN film (25 nm) with Ti sublayer (25 nm). Details of technological operations can be found elsewhere [30-33].

Fig. 2a shows typical current-voltage characteristics of the Au/Zr/oxide/TiN/Ti nanostructures with different composition and thickness of oxide films, which demonstrate reproducible bipolar switching between the low-resistance state (LRS) and the high-resistance state (HRS) after electroforming performed by applying a negative bias to the top electrode without a current compliance [30-33]. It should be mentioned that the developed nanostructures demonstrate fast switching (as low as 100 ns) and high enough endurance to the repeated programming cycles ($10^5$ more) [20].

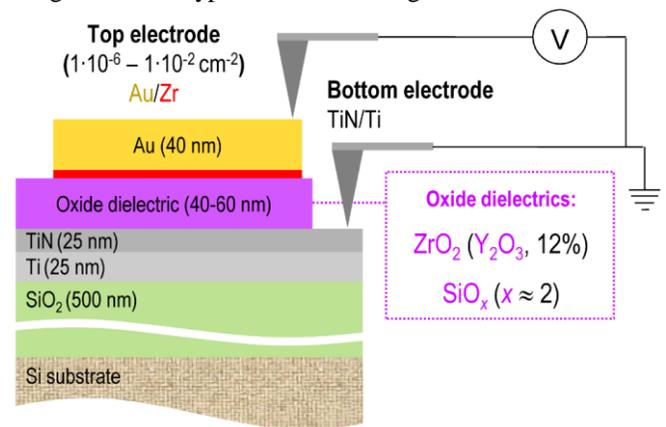

Fig. 1. Schematic cross-section of the Au/Zr/oxide/TiN/Ti memristive nanostructure.

The observed resistive switching is caused by the partial oxidation and recovery of conducting channels (filaments) in the oxide film under application of voltage of various polarity [20]. Corresponding current changes are indicated in Fig. 2a as the RESET and SET transitions. During the SET process with no current compliance, the resistance changed by 1-3 orders of magnitude, e.g. in a wide range corresponding to the extreme resistive states of the material. There is a gradual stepwise increase in current with voltage sweeping, therefore, it is possible to get a spectrum of intermediate resistive states by limiting the sweeping amplitude in the SET transition region [33]. In the RESET process, there are either abrupt or gradual current changes for given nanostructures, which, however, are poorly reproduced from cycle to cycle due to the stochastic nature of RESET process in a multifilament system [32].

The degree of resistive switching (change in resistance value) depends on the parameters of electrical stimulation and determines the adaptive behavior of memristive nanostructures [35]. The dependencies of resistance on the amplitude of negative programming voltage pulse with the duration of 5 ms are shown in Fig. 2b for the Au/Zr/oxide/TiN/Ti memristive nanostructures on the basis of different oxide films. The resistance value was measured by applying positive voltage pulse with the amplitude of 0.5 V. Current-voltage characteristics shown in Fig. 2a illustrates that the amplitude of such reading pulse must be lower than 2 V in order to preserve the current state of memristive device. It follows from Fig. 2b that the device resistance gradually decreases with the amplitude of programming pulse. It should be also noted that, after application of each programming pulse, the device was switched to its initial high resistance state by applying rather long (2 s) positive voltage pulse with the amplitude of 7 V. Such programming technique permits to control device switching to a desired resistive state in any direction and for any sequence of the resistance change.

Despite the qualitatively common behavior of resistance for different types of oxide materials, there are certain differences in the parameters of resistive switching related to the characteristics of the oxide films used. In particular, the differences are in the dynamic range of resistance and the voltage range corresponding to extreme resistance values. For example, in the case of $ZrO_2(Y)$-based memristive device, the HRS resistance is $10^5$-$10^6$ $\Omega$, but, in the case of $SiO_2$-based one, the resistance is of order $10^3$ $\Omega$, what is obviously determined by the dielectric properties of oxide layer. The values of LRS resistance (~ $10^2$ $\Omega$ for $SiO_2$ and ~ $10^3$ $\Omega$ for $ZrO_2(Y)$) are also different due to the different origin of the high-conductive states (different composition and structure of filaments) [20]. The $ZrO_2(Y)$-based devices are characterized by more abrupt (threshold-like) character of the resistance dependence: significant changes occur in the range of amplitudes of 3-4 V, whereas for the $SiO_2$-based devices the resistance changes in the wider voltage range from 2 to 5 V.

Thus, by selecting the amplitude of negative pulse it is possible to program the resistance of memristive device considered as a weight coefficient of ANN in a wide range, which depends on the oxide material. The change in resistive state of memristive device is induced by a short voltage pulse with the amplitude higher than 2 V. Accordingly, the range of ANN operating voltages, which do not cause a weight change, can be selected as ±1 V. Application of positive pulse with the amplitude up to 8 V provides the transition of memristive device to HRS. Due to the mentioned stochastic nature of resistive switching, there is a pulse-to-pulse variation of the programmed resistive value of at least 15% that should be taken into account in the development o ANN prototype.

### III. NEURAL NETWORK CONSTRUCTION PRINCIPLE

An artificial neuron as a basic element of ANN performs a weighted summation of the input signals:

$$S = \theta + \sum_{k=1}^{K} w_k x_k, \quad (1)$$

where $x_k$, $k = 1, 2, ..., K$ – inputs, $w_k$ – weight coefficients, $\theta$ – adjusting parameter, with the subsequent nonlinear transformation:

$$y = F[S]. \quad (2)$$

Nonlinearity of the neuron in (2) is usually set as a sigmoid-like function. Artificial neuron as the single-layer perceptron can be used to solve the problems of linear classification. In order to solve the problems of nonlinear classification, approximation and others, the multilayer perceptron containing at least one hidden layer should be used.

A matrix of the ANN weight coefficients is implemented in the form of a test chip with the micro-scaled Au/Zr/oxide/TiN/Ti memristive devices as shown in Fig. 3.

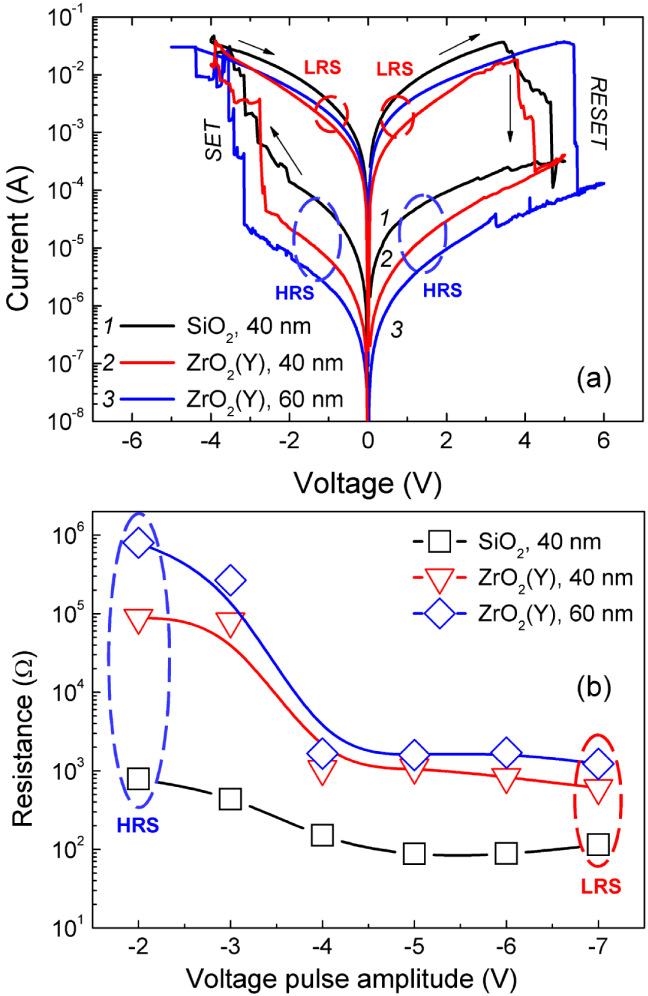

Fig. 2. Typical current-voltage characteristics of the Au/Zr/oxide/TiN/Ti memristive nanostructures based on different oxide films measured in continuous sweep mode (a) and the dependencies of resistance of the same structures on the amplitude of single voltage pulses with duration of 5 ms (b).

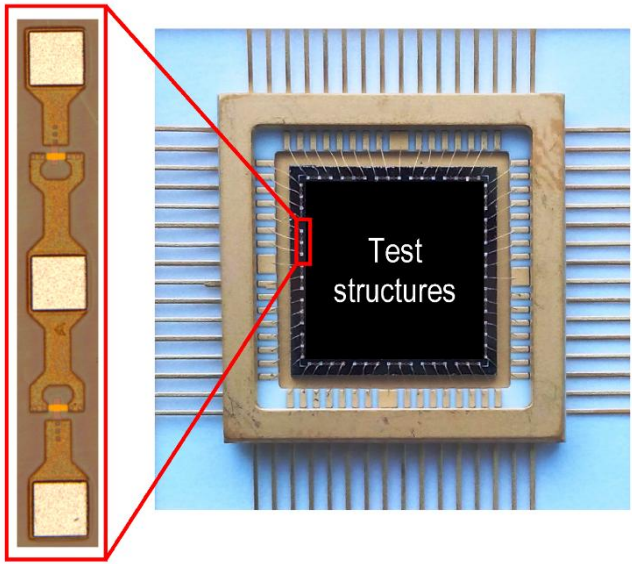

Fig. 3. The image of a test chip with the array of memristive devices for the ANN hardware implementation. The inset shows a microscopic image of the complementary pair of peripheral memristive devices.

The test chip topology contains various test structures and peripheral cross-point memristive devices mounted in the standard 64-pin metal-ceramic package intended for the hardware implementation of a double-layer perceptron (Fig. 4a). The weight matrix includes 16 pairs of memristive devices – each pair with an output in the mid-point. Such configuration is used to produce bipolar weights $w_k$ in the ANN nodes based on the complementary junction of memristive devices [25,26]. The direct input signal is sent to one memristive device, and the inverted input signal – to the other device (Fig. 4b).

The technological variation of parameters of memristive devices (device-to-device variation) can be the most important at the development stage of ANN weight matrix fabrication technology, that is why the dependence of resistance on applied voltage should be measured individually for each memristive device. Typical dependencies are presented in Fig. 2 for the large-area memristive nanostructures, but the decrease in the area of a memristive device down to 20 μm × 20 μm (in the test chip) has no significant effect on the parameters of resistive switching [20].

Each artificial neuron ($N_1 - N_4$) is implemented on the basis of general-purpose operational amplifiers (op-amps). The input stage uses a circuit of inverting summing amplifier at the $V_3, V_4$ op-amp with the increased current driving capacity. There are certain reasons for the choice of such solution. A peculiarity of memristive devices after electroforming with no current compliance is a relatively low minimum resistance (up to hundreds of Ω). Since a pulse voltage in the range of 8-10 V can be applied to the memristive device in the programming mode, the corresponding value of current through the device can reach tens of mA. The ADG436 chip is chosen as an analog multiplexer that allows loading a pulse current of up to 40 mA. The analog multiplexer provides commutation of the memristive device from the data input $D$ to the input $L$ in the programming mode of the neuron weights.

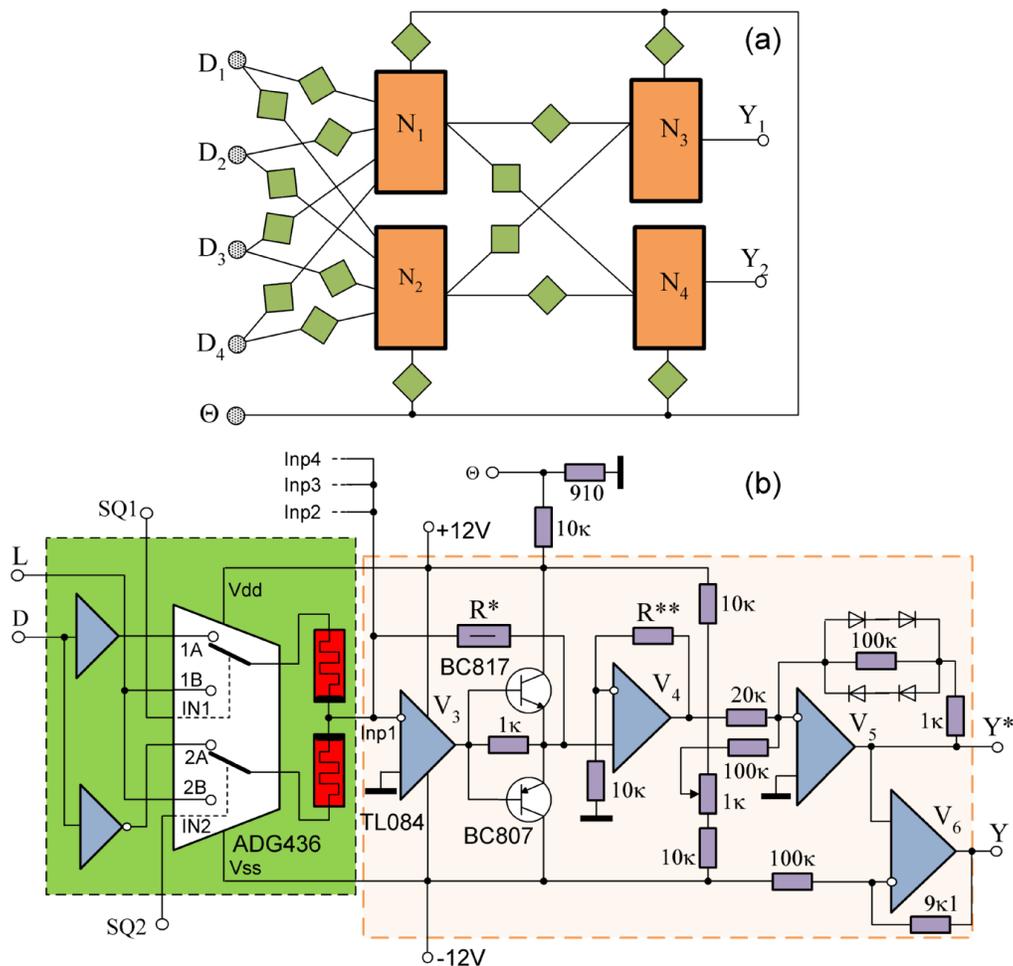

Fig. 4. Block diagram of the ANN based on complementary pairs of memristive devices (each pair is shown as a green box) (a) and the node simplified schematic circuit (b).

The inverting amplifier supports "virtual ground" at the inverting input, which simplifies the memristor switching circuitry and minimizes the quantity of multiplexers required. The resistor in the feedback of the $V_3$ op-amp is selected on the basis of a minimum resistance value of the memristive devices. In the implemented scheme, this value is chosen to be 300 Ω. Thus, each neuron forms a voltage on the summing amplifier output:

$$U_\Sigma = U_\theta \left( \frac{1}{R_1} - \frac{1}{R_2} \right) R^* + \sum_i \left( U_i \left( \frac{1}{R_{1i}} - \frac{1}{R_{2i}} \right) R^* \right), \quad (3)$$

where $U_\theta$ is the voltage at the adjusting input $\theta$ (chosen to be about 1 V), $U_i$ – the voltage at the neurons inputs, $R_{1i}, R_{2i}$ – the resistances of the upper and lower memristors according to the scheme in Fig. 4b.

The amplifier at the $V_4$ op-amp provides the required gain for summation signal in the ANN operation mode. The nonlinear transformation function $F[\cdot]$ of the neuron is implemented on the inverting voltage limiting amplifier at the $V_5$ op-amp. The output signal is formed at the output $Y^*$, and the $V_6$ op-amp interfaces the ANN output signals to the input range of analog-to-digital converters of a single-supply microcontroller (Atmel AVR-microcontroller). The microcontroller provides the ANN programming mode and interface with PC via the RS-232 protocol. Control electronics also includes a set of decoders, which form the selection signals for the current programmable memristor $SQ_n$, and the data storage registers. The input data signals $D$ and pulse voltage at the input $L$ in the memristor programming mode (training of the ANN weights) are formed according to PC commands by the microcontroller using analog-to-digital converters.

The fabricated chip with array of memristive devices and the described electronic circuits are arranged in a mobile one-board analog-digital platform, the photograph of which is presented in Fig. 5.

In the mode of weight coefficients programming, two pulses of different voltage polarity with duration of 5 ms are sequentially formed at the input $L$ of ANN: the first pulse of positive polarity switches the memristive device to HRS, the second pulse of negative polarity sets the desired resistance of memristive device. The amplitude of such programming pulse is calculated based on the dependence of resistance on applied voltage (Fig. 2b) preliminary measured for each memristive device to overcome the device-to-device variation problem. The effect of pulse-to-pulse variation of programmed weights was analyzed by measuring the variation of neuron output signals and was found to be not critical for the successful training procedure.

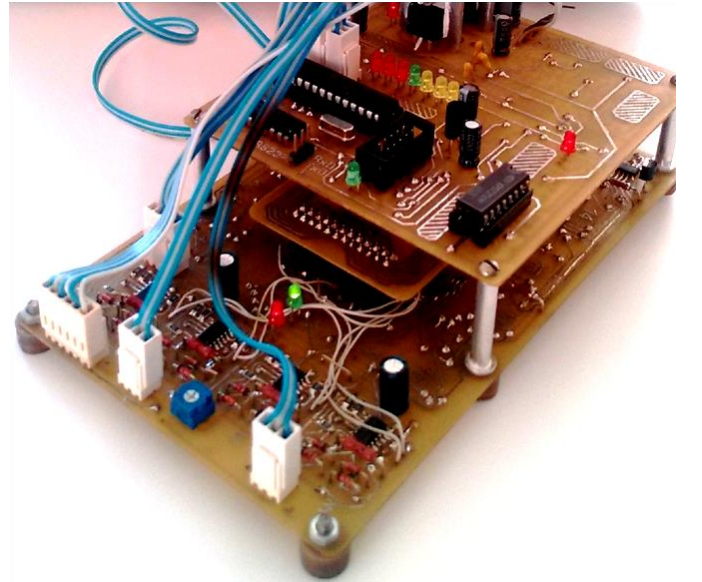

Fig. 5. The photograph of fabricated ANN board with analog (bottom) and digital (top) electronic circuits.

## IV. Neural network training

The possibility of training of the developed ANN based on memristive devices was tested on the ANN computer model. The training algorithm was developed using the error back propagation method [36] adapted to the nodes with memristive connections. As already mentioned, the dependencies of resistance on applied voltage (Fig. 2b) are measured independently for each memristive device and stored in the corresponding table. The nonlinear transformation function $F[\cdot]$ is taken the same for all neurons of the network and stored in the form of data table. The error functional $E = E(\vec{P})$ is optimized by varying vector $\vec{P} = \{W, \Theta\}^T$ containing the ANN weights $W$ and thresholds θ. At the each step of training algorithm, the weight coefficients are adjusted in the opposite direction of gradient of $E$:

$$\Delta \vec{P} = -\varepsilon \nabla E(\vec{P}), \quad (4)$$

where the coefficient $\varepsilon$ determines the step size of the algorithm and is selected from the condition $\varepsilon < 1$.

The principle of the method is to interpret $E$ as a composite function and sequentially estimate the partial derivatives as for the composite function. The sum of squared errors is used as a target function:

$$E = \frac{1}{2} \sum_j \sum_p \left( y_j^p - d_j^p \right)^2. \quad (5)$$

The input signals (samples) applied to the network are numbered by index $p$, $d$ denotes the corresponding required responses, indices $i, j$ denote the number of input point and the node number in the layer respectively, $y$ – output voltage of perceptron (result of direct propagation for the given sample). If we write (4) as the derivative of a composite function, substitute the derivative of the error (5) and introduce the notations for the delta-multipliers of output layer:

$$\delta_j = \frac{dF(S_j)}{dS_j}(d_j - y_j) \qquad (6)$$

and for the hidden layers:

$$\delta_{jl} = \frac{dF(S_{jl})}{dS_{jl}} \sum_k \delta_{k(l+1)} w_{jk(l+1)}, \qquad (7)$$

then we get a simple expression for the recursive procedure of weight coefficients correction:

$$\Delta w_{ijl} = \varepsilon \delta_{jl} x_{ijl}. \qquad (8)$$

The index $l$ is the number of ANN layer, the summation on $k$ in (7) is the summation on the nodes of the next (top) layer, which is connected to the output of the given node; respectively $w_{jk(l+1)}$ is the weight coefficient for the connection of $j$-node in the given layer to $k$-node in the next layer. For the estimation of corrections to thresholds, they may be regarded as weight coefficients for the connections with constant input $x = 1$, then:

$$\Delta \theta_{jl} = \varepsilon \delta_{jl}. \qquad (9)$$

The expression (7) contains a derivative of the activation function $F[\cdot]$. Since the activation function is tabulated, its derivative is tabulated too. The method of limitation of weight coefficients is added to the training algorithm to take into account the finite resistance range of memristive devices.

The developed double-layer perceptron was trained as a classifier of the shape of input signal (on the example of convex and concave functions) by four input points. The sampling for training was built from the following types of sequences:

$d_q = \{1 + \eta n(4q); -1 + \eta n(4q+1); -1 + \eta n(4q+2); 1 + \eta n(4q+3)\}$ – samples of concave function;
$d_q = \{-1 + \eta n(4q); 1 + \eta n(4q+1); 1 + \eta n(4q+2); -1 + \eta n(4q+3)\}$ – samples of convex function. Here $\eta = 0.25$ is the arbitrary noise level, $n(q)$ – samples of pseudo-random sequence uniformly distributed in the range [-1; 1]. The weight coefficients of the network were initialized by small random values, whereupon a series of steps of training algorithm was carried out. Fig. 6 shows typical plots of normalized error function $E$ vs. the number of training steps $M$ for different types of memristive devices. Since the error function decreases quickly, we can conclude that the learning of ANN model goes successfully.

The weight coefficients (resistance values of memristive pairs) obtained at this stage of training are considered as the initial approximation for the second phase – direct hardware

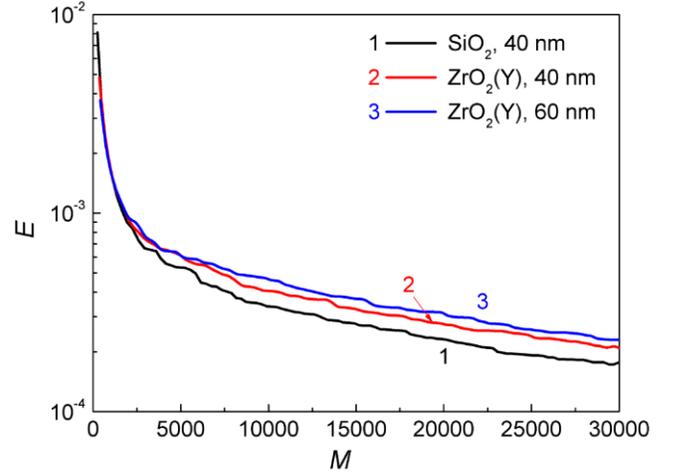

Fig. 6. The dependencies of the normalized error function on the number of training steps for the weight matrices on the basis of different types of memristive devices.

implementation of the ANN prototype with application of the described CMOS electronics.

V. CONCLUSIONS

A prototype of double-layer perceptron using an array of ZrO$_2$(Y) or SiO$_2$ based memristive devices to implement the matrix of weight coefficients is proposed. The experimentally implemented ANN circuit demonstrates the possibility to develop hybrid electronic devices on the basis of memristors and conventional CMOS electronics. The algorithm of weight coefficients training (ANN learning) is developed using the error back propagation method and takes into account finite resistance range and possible technological (device-to-device) variations of parameters of memristive devices. Preliminary experiments have shown that the ANN is also robust to the observed 15% pulse-to-pulse variations of resistance values in the weight coefficients programming mode. It is demonstrated on the computer model that the proposed ANN is capable of solving nonlinear classification problems for input signals.

The functionality of developed ANN is limited by the size of memristor-based weight matrix. However, its architecture can be extended by simple technological means. In particular, changes in topology of memristive array or its multiplication can significantly increase the number of neurons and connections between them to face the challenges of a higher functional level as a part of adaptive neurointerface for the real-time registration / stimulation of bioelectric activity of living neuronal networks in vitro (e.g. brain-on-chip neuronal cultures).